# Collaboration Success Factors in an Online Music Community

Fabio Calefato          Giuseppe Iaffaldano          Filippo Lanubile
University of Bari, Italy
{fabio.calefato, giuseppe.iaffaldano, filippo.lanubile}@uniba.it

**ABSTRACT**
Online communities have been able to develop large, open-source software (OSS) projects like Linux and Firefox throughout the successful collaborations carried out by their members over the Internet. However, online communities also involve creative arts domains such as animation, video games, and music. Despite their growing popularity, the factors that lead to successful collaborations in these communities are not entirely understood.

In this paper, we present a study on creative collaboration in a music community where authors write songs together by 'overdubbing', that is, by mixing a new track with an existing audio recording. We analyzed the relationship between song- and author-related measures and the likelihood of a song being overdubbed. We found that recent songs, as well as songs with many reactions, are more likely to be overdubbed; authors with a high status in the community and a recognizable identity write songs that the community tends to build upon.

**ACM Classification Keywords**
H.5.3. Information interfaces and presentation (e.g., HCI): Group and Organizational Interfaces – *collaborative computing, computer-supported cooperative work*.

**INTRODUCTION**
By almost exclusively relying on computer-mediated communication, participants in online communities have been able to develop large, open-source software (OSS) projects like Linux and Firefox. Besides, online communities also involve other kinds of peer productions [2] such as creative collaborations in domains like animation, video games, and music.

Despite the growing popularity of online communities, the factors that lead to the success of collaborations between members are not entirely understood [19]. For instance, we ignore whether success factors are domain dependent [15]. Previous research in OSS communities has established that successful collaborations between developers depend on both social and technical factors [8,10,21]. Also in the case of online creative arts communities, Luther et al. [14,15] found that participants' social reputation is key to the successful completion of collaborative animation efforts. Burke and Settles [3] found that users, especially newcomers, who engage in social features and one-to-one collaborations perform their songwriting goals better than those who are non-social. Other studies, instead, focused on factors that lead members of arts communities to select specific creative artifacts shared by others for reworking and recombining them into something new [5,13,20].

To further our understanding of the factors influencing the success of online creative collaborations and how they transfer across domains, we designed a study on Songtree,[1] an online community for collaborative music creation. We focused on the creative action of *overdubbing,* whereby exactly one new track is mixed with an existing audio recording (e.g., recording voice over an instrumental song), thus allowing a song to 'grow.' We analyzed the relationship between song- and author-related measures (e.g., likes, followers) and the 'success' of a song, represented by whether it is overdubbed or not. We found that recent songs, as well as songs with many reactions, are more likely to be further extended. Furthermore, authors with a high status in the community and a recognizable profile write songs that the community tends to build upon. These findings provide evidence that there are: (i) success factors specific to creative communities; (ii) common factors that are key to successful collaboration in both OSS projects and online creative arts communities.

**SONGTREE**
Born in late 2011 as a spin-off of the n-Track Software, Songtree is both an app, providing an all-in-one solution for recording songs, and an online creative music community, where artists collaborate to the creation of musical tracks. As of December 2016, the community counted about 26K registered users, of which ~5,300 are authors who uploaded over 26,000 songs.

---

[1] http://songtr.ee

Songtree allows any user to extend (namely, *overdub*) any publicly shared song in the community without permission. Songtree leverages the metaphor of a growing tree to represent and keep track of the collaborative creation of music tracks (see Figure 1). A new song is the root of the tree (the topmost node). For each song derived from it, a new branch is created and added to the song tree. Thus, over time, the tree of a song gradually grows as new overdubs are posted, each derived from any of the songs in the tree.

Several social-networking features are available in Songtree, including the ability to follow other musicians, as well as to like and bookmark songs. Analogously to Stack Overflow, Songtree also uses badges [4], which are earned by members through their activity within the community. There are three badge categories: *new songs*, *overdubs*, and *overdubs received*. Badges in these categories can be earned, respectively, by uploading new own songs, overdubbing other songs, or being overdubbed by others. Unlike Stack Overflow, however, earning badges does not unlock privileges (e.g., moderation). Instead, badges in Songtree act as a proxy measure of artists' reputation within the community, measured by the quantity and quality of the content created therein.

Finally, user profile pages (see Figure 2) list authors' personal information, biography, pictures, links to uploaded songs and followers, and statistics of their activity.

## COLLABORATION IN ONLINE COMMUNITIES

In this section, we review the literature on creative communities, focusing specifically on arts and OSS communities. For each work, we highlight the type of collaboration studied, the definition of a successful collaboration, and the relevant success factors identified.

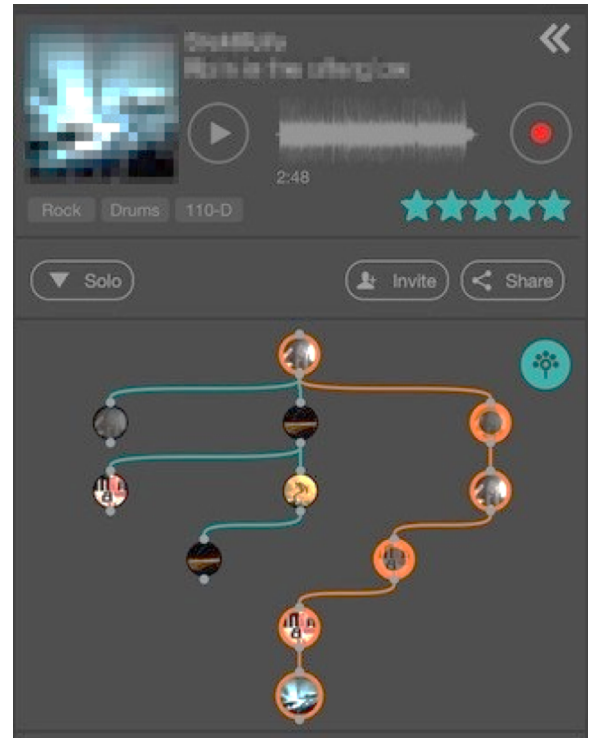

**Figure 1. The tree of a song (topmost node) with branches generated by the overdubs received.**

### Online Creative Arts Communities

Collaboration in creative arts communities typically happens in the form of *reuse*, that is, the generation of derivative content through the reworking and recombination of existing member contributions [5,13]. In music communities, reuse is mostly referred to as *remix*, where it indicates "a reinterpretation of a pre-existing song" [18].

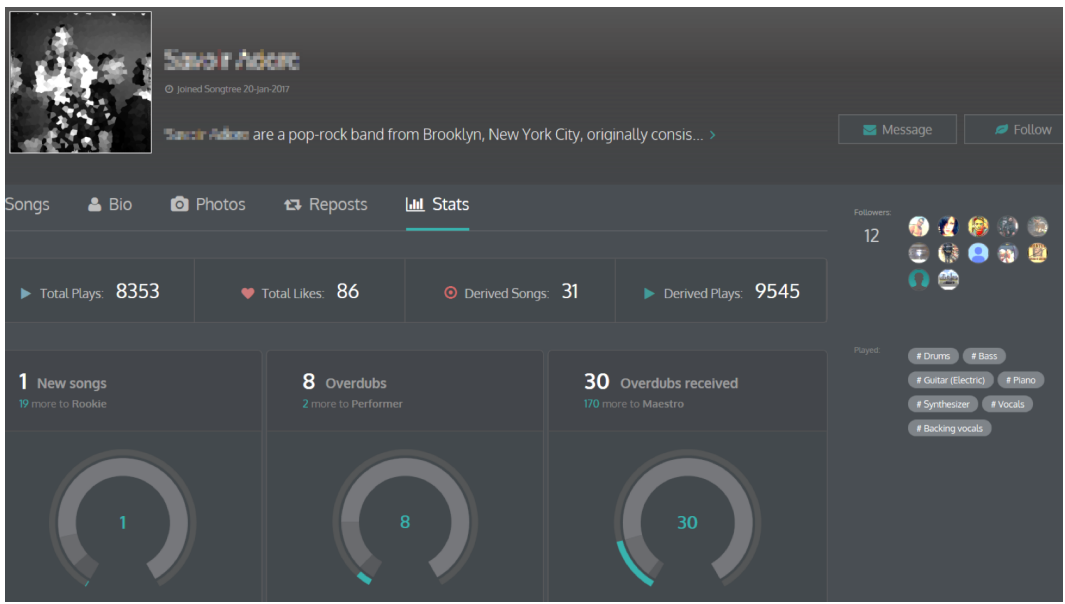

**Figure 2. The stats available in the profile page of an artist in Songtree.**

Nevertheless, the term remix has now become commonplace in creative contexts other than music to refer, for example, to reusing video animations [13] and 3D-printable content [20].

Cheliotis et al. [5] studied the likelihood of songs being remixed in the ccMixter music community. They fit a logistic regression model and observed that the positive antecedents of remixes are the following: remixes 'closer' to their parent songs (i.e., *degree of derivativity*), being an author with a history of remixes received (i.e., *fecundity*), and having a high level of commitment and contribution to the community (i.e., *social embeddedness*).

Hill and Monroy-Hernández [13] performed a study on Scratch, an online community where amateur creators combine images, music, and sound to obtain Adobe Flash-like video animations. The results of their logistic regression analysis show that the likelihood of engendering derivative works is related to work complexity, prominence of authors, and work cumulativeness. In particular, the result about cumulativeness (i.e., remixes themselves are reused more than *de novo* content) is in direct contrast with the finding by Cheliotis et al. [5] on the degree of derivativity (i.e., the 'newer' the content, the higher the likelihood of remix).

Stanko [20] investigated why some 3D-printable objects in the Thingiverse community are more generative than others. They found that the remixing likelihood is positively related to the interaction with other community members and the availability of content whose reuse would foster desired learning.

The review of prior work on remixing, especially in the music context, highlights commonalities and differences with overdubbing. Both remixes and overdubs leverage existing content to build something new. However, while remixing is 'derivative' by nature, overdubbing is 'additive,' since it implies adding a new track to an existing recording.

A few prior studies have investigated collaboration success factors in creative arts communities. Luther at al. [14,15] examined the role of leadership and other success factors in Newgrounds, a collaborative animation community. Specifically, they analyzed a dataset comprising completed (i.e., successful) and incomplete/abandoned (i.e., unsuccessful) animations created collaboratively by up to 10 co-authors. Successful collaborations are then hosted on the website, and the number of views and likes received contribute to building up authors' reputation within the community. Through a mix of qualitative and quantitative analysis, Luther at al. found that the collaborations more likely to succeed are those initiated by experienced 'leaders,' already well-known to the community, who are also inclined to communicate frequently; in addition, the quality of the idea for the animation initially presented is also positively associated with the chance of success.

Settles and Dow [19] analyzed FAWM, a hub for songwriters who meet online annually on February and collaborate to the creation of an entire album of 14 songs written in 28 days. They developed a logistic regression model on longitudinal data collected from four FAWM events (2009-2012) to predict whether any pair of songwriters would start and successfully complete a collaboration or not. Their findings, further corroborated by a survey administered to FAWM members, showed that prior interactions (i.e., exchange of direct messages) and slight differences in status are key factors in pairing, and that the perception of balanced efforts from both parties is the factor that contributes the most to the completion of such collaborations.

Overall, the review of prior work on creative arts communities highlights the existence of both content- and author-related success factors, suggesting the need for conducting the analysis of successful collaboration in Songtree on the same two levels, i.e., *songs* and *authors*.

**Open-Source Software Communities**
There has been a substantial amount of previous research on the factors affecting success in collaboration within OSS communities, where success is defined in terms of the acceptance of external code contributions (e.g., patches, pull requests) to project repositories. Baysal et al. [1] found that only about a half of the submitted pull requests to the Chrome and Firefox projects made it into the repository; Gousios et al. [9] found that about 80% of the ~170k pull requests analyzed in their study on GitHub were merged.

Regarding the factors that influence the acceptance of contributions, previous research found them to be both social and technical in nature [11,21]. Weißgerber et al. [22] analyzed successful collaboration in the form of the patch submission process in two OSS projects and found that small patches (i.e., affecting a few lines of codes and files) have significantly higher chances of being successfully merged. Yet, with the raise of 'transparent' social coding platforms such as Bitbucket and GitHub, integrators have started to make inferences about the quality of contributions not only by looking at their technical quality but also using developers' track record (e.g., history of previous contributions accepted) and reputation (e.g., number of stars and followers in GitHub) as auxiliary indicators [6,16]. Ducheneaut [7] found that contributions coming from submitters who are known to the core development team have higher chances of being accepted, as core developers also use the record of interactions as signals for judging the quality of proposed changes.

Tsay et al. [21] conducted a study where they developed a regression model to examine the association of several social and technical factors with the likelihood of accepting pull requests in GitHub projects. They found that submitters' reputation and their social distance from the code development team (i.e., track record) is surprisingly more relevant to the contribution acceptance than technical factors (e.g., the inclusion of test cases). Finally, Gousios et al. [9] consistently found in another study on GitHub that only ~13% of the pull requests reviewed were rejected for purely technical reasons.

Finally, Luther et al. [15] contrasted success factors identified in a creative community of animations (Newgrounds) to those reported in OSS literature. They reported frequent communication and project leader's experience and reputation as common factors. Our work aims at furthering this comparison.

**HYPOTHESES**

Previous work on OSS and creative arts communities suggests several potential factors, both social and technical, which may influence the success of collaboration in Songtree. In the following, we discuss a set of hypotheses, designed on two different levels of analysis, i.e., songs and authors. These hypotheses, built upon prior work, have also been revised during a couple of sessions conducted with Songtree administrators.

**Song-related hypotheses**

Community members show appreciation towards songs by liking and bookmarking them, and, of course, by playing them. Songs receiving a high amount of reactions from the community may also be showcased in Songtree under the *Popular* section of the website. As such, the visibility gained by being featured in the *Popular* section may further increase their chances to be overdubbed.

*H1: Songs that generate a high amount of reactions are more likely to be overdubbed.*

Besides, to favorite the serendipitous discovery of new artists and songs, the Songtree website also features a *Latest* section where newly uploaded contributions are promoted for also gaining visibility. Therefore, we expect that recent songs have more odds to be overdubbed.

*H2: More recently uploaded songs are more likely to be overdubbed.*

The main purpose of Songtree is to help musicians write songs collaboratively through overdubs. The more overdubs a song receives, the closer it gets to be considered 'finished.' Songtree even allows musicians to tag their songs as complete, to indicate that they do not intend to work on them anymore (albeit overdubbing may remain allowed). This speculation is in line with the finding by Cheliotis et al. [5] about the inverse relationship between a song's degree of derivativity and the chance of being overdubbed. Thus, we expect that these consolidated, more mature songs receive fewer overdubs than the others.

*H3: More mature songs are less likely to be overdubbed.*

**Author-related hypotheses**

Hill and Monroy-Hernández [13] and Cheliotis et al. [4] found that authors' prominence and their social embeddedness in the community increase the likelihood of remix. Therefore, we expect that the same applies also to Songtree.

*H4: Songs by authors with a high status in the community are more likely to be overdubbed.*

Previous research on collaboration in OSS communities has found that the identity of the submitters is a very significant factor when it comes to assessing the quality of contributions [12]. Besides, Luther et al. [14,15] found that being able to browse members' history of contributions increases the chance of successful collaboration. Songtree allows artists to maintain their profile page where they upload their avatar pictures, provide a bio, share stage photos, thus creating a personal space that reflects their identity. Some members may perceive that the effort put into curating their personal space may reflect the same attention put into creating their music. As such, we hypothesize that Songtree members are more inclined to overdub songs by fellow authors who are more easily recognized.

*H5: Songs by authors with a recognizable profile are more likely to be overdubbed.*

**EMPIRICAL STUDY**

To test the hypotheses defined earlier, we built a dataset from the entire data dump of Songtree, including data from November 2011 to December 2016. Using this dataset, we fit a logistic regression model that associates both social and technical features measured in Songtree with the likelihood of a song being overdubbed. The study approach follows a similar research by Tsay et al. [21], who previously investigated the factors influencing the likelihood of pull requests acceptance in GitHub projects.

In the rest of this section, we provide a detailed description of the dataset and the measures extracted.

**Dataset**

We built a dataset consisting of authors, songs, and song trees. Information was gathered from the entire data dump of Songtree performed in December 2016.

As initial preprocessing steps, we first removed all the songs uploaded prior to March 2015, i.e., before the song's catalog and community reached a mass large enough to not require administrators' participation anymore. Then, we filtered out all the new songs uploaded in the last 27 days. Instead, overdubs were only accounted for when they derived songs outside such time window; otherwise, they were also ditched. The 27-day threshold corresponds to the $90^{th}$ percentile of the overdub time intervals between the upload of a song and that of its first overdub. In other words, 90% of the songs in our rather sparse dataset receive their first overdub within 27 days since their upload. This step was necessary to avoid censoring issues and ensure that all the selected songs have had sufficient time to receive at least one overdub. We further discuss this preprocessing step in the Limitations section.

|  | Original dump* | Final dataset** |
|---|---|---|
| Users | 26,727 | - |
| Authors | 5,306 | 3,790 |
| Administrators | 9 | - |
| Songs | 26,055 | 16,998 |
| New songs | 15,826 | 10,595 |
| Contest songs | 136 | - |
| Closed songs | 39 | - |
| Hidden songs | 5,063 | - |
| Orphan songs | 207 | - |
| Self-overdubs | 3,843 | - |
| Overdubs | 10,229 | 6,403 |
| Remixes | 52 | - |

*as of Dec. 2016; ** Mar. 2105 – Nov. 2016*

**Table 1. A comparison of data in the Songtree dump and the final dataset (rows in grey are filtered out).**

A breakdown of the data dump is reported in Table 1. As of December 2016, the community counted about 26,000 members, of which ~5,300 (20%) are authors who had recorded and shared at least one on Songtree. Overall, over 26,000 songs had been upload to Songtree, of which ~15,000 new songs and ~10,000 overdubs.

To build the final dataset, we further preprocessed the dump and excluded the content matching the following criteria:

- *Non-authors* – community members who have not recorded and shared any song on Songtree. They are excluded because they have gained no reputation as authors.
- *Administrators* – accounts registered by the members of the Songtree development team. We opted for excluding administrators' accounts and the content shared by them (e.g., contest songs) to avoid altering our findings on how the Songtree community behave when collaborating.
- *Closed songs* – songs that, as per author's setting, either cannot be overdubbed or are marked as finished. Hence, they are excluded because, respectively, they disable or discourage overdubbing. Note that these are just leaf nodes. Instead, the song trees they belong to are retained.
- *Hidden songs* – songs that, as per author's setting, are not publicly listed and can be found and overdubbed only if the author shares a link with others. These songs are used by authors who want their music to remain private or keep the collaboration restricted to their inner circle.
- *Orphan songs* – songs that belong to no song tree and overdubs derived from no parent songs.
- *Self-overdubs* – any child song derived from a parent song recorded by the same author. They are excluded because they do not represent meaningful cases of collaborative songwriting.
- *Remixes* – songs that only add effects or alter frequencies through the equalizer. They are excluded because they do not help their authors earn badges or improve their status.
- *Contest songs* – songs uploaded by the Songtree team to start contests with prizes awarded to the best overdubs.

At the end of the preprocessing stage, we ended up with a final dataset consisting of 16,998 songs (10,595 new songs + 6,403 overdubs), and 3,790 authors.

**Measures**

From the final dataset built, we defined several measures to inform our analysis (see Table 2). Since the outcome measure for our statistical model is whether a song has been overdubbed, we define `overdubbed` as a dichotomous, yes/no variable.

To draw causal conclusions, all the measures detailed next are calculated prior to the event '*song has received the first overdub.*' As an example, consider a song *S* from our dataset. First, in case that *S* has been overdubbed, we retrieve the time $T_S$ when it received the first overdub. Then, all the measures are calculated at the instant before $T_S$ (e.g., we compute the number of likes received by *S before S* was first overdubbed). We can safely ignore any subsequent overdub of *S* because we conduct a logistic regression and our outcome variable is binary (i.e., *S* has received 0 overdubs vs. *S* has received 1+ overdubs). For songs that received no overdubs, instead, we retrieve the timestamp of the last event recorded in the entire dump and, hence, we apply no time-based measurement.

In the following, the measures are described accordingly to the two levels of analysis, that is, songs and authors.

*Song-level measures*

For our song-related measures, we capture various dimensions as well as signals expressed by community members in form of likes and plays. These measures are presented in the following, grouped by hypothesis/main factor.

*H1 – Reactions*

`#likes_received` – A numeric variable that counts how many times the song has been liked by someone in Songtree.

`#bookmarks_received` – A numeric variable that counts how many times the song has been bookmarked by someone in Songtree.

`#times_played` – A counter of the number of times that a song has been played in Songtree.

|  | Measure | Type | Description | Hypothesis |
|---|---|---|---|---|
|  | overdubbed | nominal | Whether the song has received any overdubs or not. Values: {*Yes, No*} | - |
| Song level | #likes_received | ratio | No. of likes that the song received by Songtree members. | H1 |
| Song level | #bookmarks_received | ratio | The number of times that the song has been bookmarked. | H1 |
| Song level | #times_played | ratio | No. of times the song was played in Songtree. | H1 |
| Song level | overdub_time_interval | interval | Time difference (in minutes) between the respective upload times on Songtree of an overdub (if any) and its parent song. | H2 |
| Song level | song_depth | ratio | The distance in number of nodes from the root song that started the song tree. It is 0 for root songs. | H3 |
| Author level | #followers | ratio | No. of users following author's activities on Songtree. | H4 |
| Author level | songtree_ranking | ratio | $\frac{\#followers + \#user\_likes + \#user\_plays + \#derived\_plays}{\#shared\_songs}$ | H4 |
| Author level | new_songs_badge | ordinal | Badge gained by uploading new songs. Values: {*None, Rookie, Songwriter, Composer*} | H4 |
| Author level | overdubs_badge | ordinal | Badge gained by overdubbing other authors' songs Values: {*None, Performer, Top performer, Virtuoso*} | H4 |
| Author level | overdubs_received_badge | ordinal | Badge awarded when enough overdubs are recorded an authors' songs. Values: {*None, Songsmith, Band leader, Maestro*} | H4 |
| Author level | has_avatar | nominal | Whether the author has uploaded a profile picture or not. Values: {*Yes, No*} | H5 |

**Table 2. Measures defined for the study, grouped by level of analysis and hypothesis.**

*H2 – Recent Songs*

overdub_time_interval – A variable measuring (in seconds) the time interval between the upload of an overdub and that of the parent song from which it was derived.

*H3 – Mature Songs*

song_depth – A continuous variable measuring the length of the path to its root. A root song (i.e., a new song) has depth 0.

*Author-level measures*

Regarding the author-related measures, we capture various signals and dimensions of identity, social status, and productivity within the community.

*H4 – Status in Community*

#followers – A numerical variable that counts the number of followers of an author.

songtree_ranking – A *coolness* index, updated on a weekly basis and used by Songtree administrators to rank the community authors. It is computed per author as follows:

$$\frac{\#followers + \#user\_likes + \#user\_plays + \#derived\_plays}{\#shared\_songs}$$

where #followers is the same measure defined earlier, #user_likes is the cumulative number of likes received by all the songs by the author, #user_plays is the cumulative number of times that all the songs by the author have been played, #derived_plays is the cumulative number of times that all the songs derived from the author's songs have been played, and #shared_songs is the number of tracks shared by the author on Songtree.

new_songs_badges – Depending on the number of new songs uploaded, users unlock badges that reflect their level of productivity. This ordinal variable is defined in the set {*None, Rookie, Songwriter, Composer*}. The *Rookie* badge is unlocked by uploading at least 20 new songs. The *Songwriter* badge is unlocked by uploading at least 50 new songs. The last one, *Composer*, is unlocked by uploading at least 320 new songs.

overdubs_badges – Depending on the number of overdubs uploaded, users unlock badges that reflect their attitude towards contributing to others' songs. This ordinal variable is defined in the set {*None, Performer, Top performer, Virtuoso*}. The *Performer* badge is unlocked by uploading at least 10 overdubs. The *Top performer* badge is unlocked by uploading at least 40 overdubs. The last one, *Virtuoso*, is unlocked by uploading at least 200 new songs.

`overdubs_received_badges` – Depending on the number of overdubs received by their own songs, users unlock badges that reflect the extent to which their songs attract external contributions from other authors in Songtree. This ordinal variable is defined in the set {*Songsmith*, *Band leader*, *Maestro*}. The *Songsmith* badge is unlocked by uploading at least 10 overdubs. The *Band leader* badge is unlocked by uploading at least 40 overdubs. The last one, *Virtuoso*, is unlocked by uploading at least 200 new songs.

*H5 – Recognizable Profile*

`has_avatar` – A dichotomous variable indicating whether the author has uploaded a personal picture to customize the profile image.

**RESULTS**

In the following, we examine our hypotheses and how each measure (i.e., predictor variable) is associated with our dependent variable that models a successful collaboration, i.e., whether a song has been overdubbed.

Using these measures, we created a model that predicts the likelihood of a song being overdubbed (see Table 3). Specifically, we fit a multi-level mixed-effects logistic regression model to our data. We choose a logistic regression approach to predict better our binary outcome variable. The multi-level approach accounts for the two layers of the dataset, with the variable *song* nested under *author*. Finally, in the mixed-effect model, all measures are *fixed effects* except for the author, which is represented as a *random effect* term; this allows us to capture author-to-author variability in the response variable (`overdubbed`), that is, some authors write songs that are more likable or leave more room for collaboration than others.

To perform the logistic regression, we used the `lme4` R package, which accounts for cross-classification of data, as authors appear in multiple song trees in our dataset. To ensure normality, each of the continuous variables in the model was log-transformed and standardized, such that the mean of each measure is 0 and the standard deviation is 1. Furthermore, we checked our dataset for multicollinearity problems [7]. We first computed the correlation matrix for the predictors and found that `#followers` and the three badges-related measures (i.e., `new_songs_badge`, `overdubs_badge`, `overdubs_received_badge`), as well as `times_played` and `songtree_ranking`, have strong pairwise correlations (i.e., ≥ 0.7). Therefore, to fix the multicollinearity, we retained the `times_played` and `#followers` measures while discarding the others. Then, after fitting the model, we checked again for collinearity using the Variance Inflation Factor (VIF) and found no value larger than 4.

In Table 3, we report the results of the logistic model. All predictors are important, that is, statistically significant at 1% level ($p < 0.01$) or smaller, as obtained from Wald test. Given the large size of our dataset, however, we report and discuss predictor contributions in terms of odds ratio, which is an unstandardized effect size statistic that tells the direction and the strength of the relationship between predictors and the odds that a song is overdubbed, i.e., the increase or decrease of the odds of 'success' occurring per 'unit' of the measure. We remind the reader that an odds ratio close to 1 means that exposure to property A (i.e., one of the considered predictors) does not affect the odds of a song being overdubbed. An odds ratio far smaller than 1 is instead significantly associated with lower odds. Conversely, an odds ratio much larger than 1 means that there are higher odds for a song to be overdubbed with exposure to predictor A.

Finally, to evaluate the fit, in Table 3 we report *AIC* and $R^2$ for the statistical model developed [17]. Regarding $R^2$, we used the `MuMIn` package in R to computed both the marginal ($R_m^2$) and conditional ($R_c^2$) version: the former describes the proportion of variance explained by fixed effects alone, whereas the latter combines fixed and random effects together. Results show that our model fits the data very well, as it explains 95% of the variability of the data ($R_c^2 = 0.95$).

**H1 – Songs that generate a high amount of reactions are more likely to be overdubbed**

To test our first hypothesis, we examined the association between the probability of a song to receive an overdub and the number of times that the song has been liked and bookmarked.

| | Factor | Measure (predictor) | Odds Ratio |
|---|---|---|---|
| Song level | | *(Intercept)* | 0.068 |
| | Reactions (H1) | #likes_received | 1.176** |
| | | #bookmarks_received | **2.116*** |
| | Recent Songs (H2) | overdub_time_interval | **0.002*** |
| | Mature Songs (H3) | song_depth | **0.557*** |
| Author level | Status in Community (H4) | #followers | 0.817*** |
| | | songtree_ranking | **9.479*** |
| | Recognizable Profile (H5) | has_avatar (default: false) | **3.114*** |
| | | **AIC** | 3925 |
| | | $R_m^2$ | 0.94 |
| | | $R_c^2$ | 0.95 |

Table 3. The logistic mixed-effects model for the likelihood of song overdubbing (sig.: ** p<0.01, *** p<0.001).
Predictors with large effect sizes are shown in bold.

From the odds ratios reported in Table 3, we observe that while the association of the likelihood of a song being overdubbed with the measure `#likes_received` (1.176) is negligible, the `#bookmarks_received` predictor is positively and strongly associated with the likelihood of the song to be overdubbed (2.116).

Therefore, we found strong support for H1.

### H2 – More recently uploaded songs are more likely to be overdubbed

We tested H2 by measuring the time interval between the upload of an overdub and the upload of its parent song.

From the results reported in Table 3, we observe that the measure of `overdub_time_interval` is one of the strongest predictors in our model (0.002). Specifically, the small odds ratio for this predictor indicates that time is strongly and negatively associated with the odds of a song being overdubbed (i.e., the longer since the upload of the song, the smaller the chances to be overdubbed).

Therefore, we found strong support for our hypothesis H2.

### H3 – More mature songs are less likely to be overdubbed.

We tested H3 by measuring the distance of a song from the root of its song tree. We found the measure `song_depth` to be strongly and negatively correlated with the likelihood of the song to be overdubbed (0.557).

Hence, also H3 is supported.

### H4 – Songs by authors with a high status in the community are more likely to be overdubbed

To test H4, we look at the author-level predictors in our statistical model (see Table 3).

Specifically, while the effect of `#followers` is negligible (0.817), the `songtree_ranking` predictor is positively and very strongly associated with the likelihood of highly-ranked author's songs being overdubbed (9.479).

Therefore, we found support for our hypothesis that the higher the authors' status in Songtree, the higher the odds of their songs to be overdubbed.

### H5 – Songs by authors with a recognizable profile are more likely to be overdubbed

We tested H5 by examining the association of the dichotomous predictors `has_avatar` with the likelihood of songs to be overdubbed. We found that the predictor is strongly and positively associated with the outcome variable (3.114).

Therefore, we found strong support for our fifth hypothesis, according to which songs by authors whose profile is easily recognized are more likely to be overdubbed.

## DISCUSSION

In this section, we summarize the results, discuss them in terms of prior research, and show the implications for the Songtree community and developers.

### Popular Songs

The results for our first hypothesis H1 show that 'popular' songs, i.e., songs that generate larger amounts of reactions in terms of bookmarks have significantly higher chances to be overdubbed.

Regarding OSS communities, this lack of evidence is explained by the nature of code contributions, as there are no 'popular' pull requests or patches. In fact, in their analysis of pull request acceptance in GitHub, Tsay et al. [21] discuss the effect of popularity at the project level, using the number of stars and collaborators as proxies. This finding complements those previous studies performed on other creative arts communities, such as Newgrounds [14,15] and FAWM [3,19], who did not evaluate the popularity of creative artifacts as a predictor of future successful collaborations. In this regard, Songtree is already leveraging these features, as it promotes popular songs in a dedicated section of its landing page.

### Recent and Mature Songs

Regarding H2, we found very strong evidence that the longer since the upload of a song, the less its chance to be overdubbed. The cause of this strong effect arguably depends on how Songtree provides recommendations to its members, that is, by suggesting recent songs, usually uploaded in about the last two days. Given the incidence of the time factor in the current implementation of the recommender system, introducing some randomness or other factors would favorite the discovery of 'older' songs.

Regarding H3, we found that more mature songs, i.e., songs created towards the end of a long collaboration process, are less likely to be overdubbed. In fact, the predictor that measures the depth of a song in its tree is significantly and negatively associated with the likelihood of being overdubbed. This finding confirms the results reported by Cheliotis et al. [5] on the degree of derivativity. A related finding is also reported by Hill and Monroy-Hernández [13] who found that older remixes, comparable to mature songs close to completion, are reused less because believed to be more complex. Our finding is particularly relevant to musicians seeking to increase the number of overdubs received by their songs. To this end, they should either start new songs or overdub less mature songs so that others still have enough 'room' to build upon their own work.

### Status and Profile of Authors

The results our study provide support for our hypothesis H4 that songs by authors who have gained a higher status in the Songtree community are more likely to be overdubbed. This finding is consistent with previous results from previous studies on arts communities [5,13–15]. Currently, Songtree shows a rank (updated weekly) of top artists in a dedicated section of the website. To further leverage the effect of author prominence in the community, the visibility of other status signals might be increased, for example, by showing next to the author's name also the number of followers and other counters, which are now only accessible by visiting the profile page.

Furthermore, we also provide support for hypothesis H5, according to which songs by authors easily recognized by their profile picture are overdubbed more. Surprisingly, Stanko [20] found that promoting artists and songs on the front page of Thingiverse community website had no impact on the likelihood of remixing. Still, our finding confirms the evidence provided by an earlier study on OSS communities, which found that the identity of pull request submitters is a very significant factor when it comes to assessing the quality of contributions [12]. Albeit not entirely novel, our finding is particularly relevant to the Songtree community because it is another actionable success factor among those identified in the study. As such, authors looking for increasing their visibility in the community should select their avatar picture carefully and leverage other platform features that may make their identity easier to recognize.

Overall, given the results of hypotheses H4 and H5, our study provides new evidence that the status and the identity of community members are consistently used as proxies of artifact quality regardless of the domain, whether technical (as in OSS communities) or artistic (as in music communities).

**Causal relationships**
One of the main contributions of this observational study is the certainty about the direction of causality for the predictors used to build the statistical model and the outcome variable.

In fact, because all the values have been computed prior to each event '*song is overdubbed*,' the predictors are inherently robust to reverse causality. As such, we can make inferences about the underlying causal relationships uncovered by our study and state that the increase in the likelihood of song overdubbing is the result of the occurrence of any of the events and song properties described earlier.

**Limitations**
The main limitation of this work concerns the generalizability of our findings. We cannot affirm that Songtree is representative of all online music communities, nor we are certain that our findings would transfer to other types of online creative arts communities. We intend to address this limitation by replicating the study on other music communities, as well as on creative arts communities of a different genre.

Regarding the findings on the derivability of recent and mature songs (H2), we acknowledge the risk that the strength of the related predictors in the regression model is due to the recommender system adopted to feed the list of *Recent* and *Popular* songs on Songtree's website. Future replications on other communities will help us overcome also this threat.

As for dataset construction, we chose to exclude all the songs uploaded in the last 27 days to ensure that all the songs in the dataset had had sufficient time to be overdubbed at least once. Of course, the older a song, the more overdubs it may have received. Therefore, rather than considering the number of received overdubs, we consistently chose to build a logistic regression model that predicts the likelihood of a song being overdubbed (i.e., 1+ times) or not. Future works involving the development of linear regression models to predict the number of overdubs will require a finer approach with longitudinal analysis.

Another limitation concerns the validity of the construct defined to assess the extent to which authors are recognizable in Songtree. While we acknowledge that using only a dichotomous variable (i.e., `has_avatar`) to model an entire author's identity may appear simplistic, we underline that the having a customized profile picture is one of the strongest predictors in the developed statistical model.

Cheliotis et al. operationalized the prominence of authors in the community using a bow-tie analysis [5]. Instead, we relied on Songtree's own metric of commitment. In our future extension, we aim to leverage clique analysis for a finer-grain evaluation of author status as an overdub factor.

Previous research has found that communication has a positive relationship with remixing [19,20]. In our future work, we will extend our explanatory model to include message exchange.

Finally, we acknowledge that the lack of qualitative analysis (e.g., questionnaires) does not allow us to triangulate data and gather a better understanding of our findings. We intend to overcome this limitation in a future extension of the study.

**CONCLUSION**
In this paper, we examined the factors that influence the songwriting collaboration within the Songtree music community. Specifically, we created a statistical model to analyze the relationship between song- and author-related measures and the likelihood of songs being overdubbed (i.e., completed thanks to other musicians).

We found that recent songs, as well as songs with many reactions, are more likely to be derived. These success factors related to the popularity of artifacts are specific to online creative arts communities. Furthermore, authors with a high status in the community and a recognizable identity write songs that continue to be evolved by the community. Comparing to similar results in OSS communities, we have evidence that people-related factors are also key to successful contribution in OSS projects. Our findings are also useful to inform the administrators of creative arts communities about aligning their policies with the signals that members are already using to discover songs and collaborate.

As future work, we intend to improve our statistical model by including new predictors. For example, we intend to study the effect of messages exchanged by authors and their polarity. In addition, author status within Songtree can also be assessed through social network analysis measures, such as degree centrality, calculated from the collaboration and communication networks. Furthermore, as we acquire further snapshots of Songtree's database, we intend to conduct a longitudinal study to uncover new success factors, regarding, in particular, the retention and loyalty of community members over time.

**ACKNOWLEDGMENTS**

We are grateful to Songtree for opening their data for research purposes.

**REFERENCES**
1. O. Baysal, R. Holmes, and M.W. Godfrey. 2012. Mining usage data and development artifacts. In *IEEE Int'l Working Conf. on Mining Software Repositories* (MSR '12), 98–107.
2. Y. Benkler. 2007. *The wealth of networks: How social production transforms markets and freedom*. Yale University Press, New Haven, CT, USA.
3. M. Burke and B. Settles. 2011. Plugged in to the community. In *Proc. of the 5th Int'l Conf. on Communities and Technologies* (C&T '11).
4. F. Calefato, F. Lanubile, M.C. Marasciulo, and N. Novielli. 2015. Mining successful answers in stack overflow. In *IEEE Int'l Working Conf. on Mining Software Repositories*, 430–433.
5. G. Cheliotis, N. Hu, J. Yew, and J. Huang. 2014. The Antecedents of Remix. In *Proc. of the 17th ACM Conf. on Computer Supported Cooperative Work* (CSCW '14), 1011–1022.
6. L. Dabbish, C. Stuart, J. Tsay, and J. Herbsleb. 2012. Social Coding in GitHub: Transparency and Collaboration in an Open Software Repository. *Proc. of the ACM 2012 Conf. on Computer Supported Cooperative Work*: 1277–1286.
7. N. Ducheneaut. 2005. Socialization in an open source software community: A socio-technical analysis. *Computer Supported Cooperative Work: CSCW: An Int'l Journal* 14, 4: 323–368.
8. G. Gousios, M. Pinzger, and A. van Deursen. 2014. An exploratory study of the pull-based software development model. In *Proc. of the 36th Int'l Conf. on Software Engineering - ICSE 2014*.
9. G. Gousios, M. Pinzger, and A. van Deursen. 2014. An Exploratory Study of the Pull-based Software Development Model. In *Proc. of the 36th Int'l Conf. on Software Engineering*, 345–355.
10. G. Gousios, M.-A. Storey, and A. Bacchelli. 2016. Work practices and challenges in pull-based development: The contributor's perspective. In *Proc. of the 38th Int'l Conf. on Software Engineering - ICSE '16*.
11. G. Gousios, M.-A. Storey, and A. Bacchelli. 2016. Work Practices and Challenges in Pull-Based Development: The Contributor's Perspective. In *Proc. of the 38th Int'l Conf. on Software Engineering - ICSE '16*, 285–296.
12. G. Gousios, A. Zaidman, M.-A. Storey, and A. van Deursen. 2015. Work Practices and Challenges in Pull-Based Development: The Integrator's Perspective. In *2015 IEEE/ACM 37th IEEE Int'l Conf. on Software Engineering*.
13. B.M. Hill and A. Monroy-Hernández. 2012. The Remixing Dilemma. *American Behavioral Scientist* 57, 5: 643–663.
14. K. Luther and A. Bruckman. 2008. Leadership and success factors in online creative collaboration. In *IEEE Potentials* (CSCW '08), 27–32.
15. K. Luther, K. Caine, K. Ziegler, and A. Bruckman. 2010. Why it works (when it works). In *Proc. of the 16th ACM Int'l Conf. on Supporting group work* (GROUP '10).
16. J. Marlow, L. Dabbish, and J. Herbsleb. 2013. Impression Formation in Online Peer Production: Activity Traces and Personal Profiles in GitHub. *16th ACM Conf. on Computer Supported Cooperative Work*: 117–128.
17. S. Müller, J.L. Scealy, and A.H. Welsh. 2013. Model Selection in Linear Mixed Models. *Statistical Science* 28, 2: 135–167.
18. E. Navas. 2012. *Remix Theory: The Aesthetics of Sampling*.
19. B. Settles and S. Dow. 2013. Let's get together. In *Proc. of the SIGCHI Conf. on Human Factors in Computing Systems - CHI '13* (CHI '13), 2009.
20. M.A. Stanko. 2016. Toward a Theory of Remixing in Online Innovation Communities. *Information Systems Research* 27, 4: 773–791.
21. J. Tsay, L. Dabbish, and J. Herbsleb. 2014. Influence of social and technical factors for evaluating contribution in GitHub. In *36th Int'l Conf. on Software Engineering*, 356–366.
22. P. Weißgerber, D. Neu, and S. Diehl. 2008. Small patches get in! In *Proc. of the 2008 Int'l workshop on Mining software repositories* (MSR '08), 67.